\documentclass{appolb}
\usepackage{amsmath,amsfonts,amssymb}

\begin{document}
\title{Soft-wall modelling of meson spectra\thanks{Presented at Excited QCD 2016.}}
\author{S. S. Afonin\address{Saint Petersburg State University, 7/9 Universitetskaya nab., St.Petersburg, 199034, Russia}}
\maketitle

\begin{abstract}
The holographic methods inspired by the gauge/gravity
correspondence from string theory have been actively applied to
the hadron spectroscopy in the last eleven years. Within the
phenomenological bottom-up approach, the linear Regge-like
trajectories for light mesons are naturally reproduced in the
so-called "Soft-wall" holographic models. I will give a very short
review of the underlying ideas and technical aspects related to
the meson spectroscopy. A generalization of soft-wall description
of Regge trajectories to arbitrary intercept is proposed. The
problem of incorporation of the chiral symmetry breaking is
discussed.
\end{abstract}

\section{Introduction}

The hypothesis of AdS/CFT correspondence  from string
theory~\cite{mald} (also referred to as gauge/gravity duality or
holographic duality) has led to unexpected and challenging ways
for description of strongly coupled systems. Such descriptions are
given in terms of weakly-coupled higher-dimensional gravitational
theories. The holographic ideas have penetrated to many branches
of physics. I will consider a traditional (and one of the first)
application of AdS/CFT correspondence --- the hadron physics. This
field is very extensive and in the given short report I discuss
only some basic elements for description of Regge-like meson
spectrum within a phenomenological bottom-up holographic approach.

The physics of hadrons composed of light quarks is highly
non-perturba\-tive. Such hadrons represent typical strongly
coupled systems. The masses of hadron states appear in poles of
correlation functions of QCD currents interpolating these states.
The AdS/CFT correspondence provides a practical recipe for
calculation of correlation functions in strongly coupled gauge
theories via the semiclassical expansion of the action of
higher-dimensional dual theory~\cite{witten}. The correlation
functions encode various dynamical information. If we are
interested only in the mass spectrum (representing a part of this
information) it is sufficient to solve the equations of motion
using the plane-wave ansatz for physical particles. The obtained
infinite tower of Kaluza-Klein states represents a model for an
infinite number of meson resonances expected in the large-$N$
limit. In the bottom-up approach, one assumes the existence of a
dual theory for QCD, guesses an action for this theory and use it
to do all necessary calculations in the semiclassical limit. Below
some examples are demonstrated.

\section{The Soft-wall model}
\subsection{Vector mesons}

The first and mostly used in practice bottom-up holographic model
that describes the linear Regge-like meson spectrum was
constructed in Ref.~\cite{son2}. The simplest action of this model
describing the vector mesons is
\begin{equation}
\label{1a}
S=-\frac{1}{4g_5^2}\int d^4\!x\,dz\sqrt{g}\,e^{-az^2}F_{MN}F^{MN},
\end{equation}
where $F_{MN}=\partial_M V_N-\partial_N V_M$, $M=0,1,2,3,4$ (the
metric is mostly negative), and $g_5$ represents the 5D vector
coupling. The action~\eqref{1a} is defined in the 5D anti-de
Sitter (AdS$_5$) space. A widely used parametrization of its
metric is: $ds^2=\frac{R^2}{z^2}(dx_{\mu}dx^{\mu}-dz^2)$. Here $R$
is the AdS$_5$ radius and $z>0$ denotes the holographic coordinate
which has the physical sense of inverse energy scale. On the
boundary of the AdS$_5$ space, $z=\epsilon\rightarrow0$, the
vector field $V_M$ corresponds to the source for a QCD operator
interpolating the vector mesons, we will put
$V_M(x,\epsilon)\leftrightarrow\bar{q}\gamma_{\mu}q$. The AdS/CFT
correspondence provides various prescriptions for connections
between a gauge theory and its holographic dual~\cite{witten}. One
of them yields the masses of dual fields in the AdS$_5$ space,
\begin{equation}
\label{3a}
m_5^2R^2=(\Delta-J)(\Delta+J-4),
\end{equation}
where $\Delta$ is the dimension of the corresponding $J$-form
operator. In practice, $J$ means just spin. In the case under
consideration, $\Delta=3$, $J=1$, and we have $m_5=0$
in~\eqref{1a}.

The linear Regge-like meson spectrum emerges due to the background
$e^{-az^2}$ in action~\eqref{1a}. This background resembles the
dilaton coupling in some string theories. It provides a "soft" way
for introducing the mass scale $a$. The name "Soft-wall" (SW)
model is used for such actions in order to distinguish from the
"Hard-wall" holographic models proposed earlier~\cite{son1}, where
the mass scale appears via a hard cutoff at some $z_0$ and the
spectrum depends on the boundary conditions imposed at $z_0$. The
meson spectrum of Hard-wall models is not Regge-like. In the SW
models, the only boundary condition is that the action must be
finite.

The gauge invariance of action~\eqref{1a} allows to choose a
convenient gauge $V_z=0$. The physical particles correspond to the
plane-wave ansatz $V_\mu(x,z)=v(q,z)e^{iqx}\varepsilon_\mu$. The
equation of motion for the scalar function $v(q,z)$ following
from~\eqref{1a} reads,
\begin{equation}
\label{5a}
\partial_z\left(\frac{e^{-az^2}}{z}\partial_z v\right) + \frac{e^{-az^2}}{z}q^2v=0.
\end{equation}
The eigenfunctions of Eq.~\eqref{5a} satisfy $v_n(q,0)=0$ and the
discrete mass spectrum is given by the corresponding eigenvalues
$q_n^2=m_n^2$,
\begin{equation}
\label{6}
m_n^2=4|a|\left(n+1\right),\qquad n=0,1,2,\dots.
\end{equation}
The spectrum does not depend on the sign of $a$. Note, however,
that the choice $a<0$ results in emergence of a non-normalizable
massless state~\cite{son2,son3}. This mode gives a finite
contribution to the action as usual normalizable states and for
this reason leads to an unphysical massless pole in the two-point
vector correlator~\cite{son3}.

\subsection{Scalar mesons}

The action of SW model for free 5D scalar fields is
\begin{equation}
\label{31a}
S=\frac{1}{2}\int d^4\!x\,dz\sqrt{g}\,e^{-az^2}(\partial_M\Phi\partial^M\Phi-m_5^2\Phi^2).
\end{equation}
Solving the ensuing equation of motion one arrives at the spectrum
\begin{equation}
\label{33a}
m_n^2=2|a|\left(2n+1+\sqrt{4+m_5^2R^2}+\frac{a}{|a|}\right),\qquad
n=0,1,2,\dots.
\end{equation}
Here the problem with unphysical zero mode at $a<0$ does not
appear since its contribution to the action is infinite. In
addition, Eq.~\eqref{33a} suggests that if $m_5^2R^2=-4$ (the
minimal allowed value of the mass squared in the AdS$_5$
space~\cite{freedman}) one has a physical massless state. The
given observation provides a possibility to incorporate the
pseudogoldstone $\pi$-meson~\cite{br3}. According to
prescription~\eqref{3a}, such a 5D mass would correspond to the
canonical dimension $\Delta=2$ of the interpolating operator. The
gauge-invariant local operators in QCD cannot have this dimension.
One might speculate that $\Delta=2$ corresponds to the current
$\partial_\mu\pi$ in the low-energy strong interactions and that
the choice $a<0$ in the dilaton background somehow mimics the
dominance of pion background in the deep infrared
$z\rightarrow\infty$.

\subsection{Higher spin mesons}

Two different descriptions of the Higher Spin Fields (HSF) were
used in the SW models: The gauge-invariant one~\cite{son2} and a
more general description of Ref.~\cite{br4}. I will discuss the
first method.

The free massless HSF are described by symmetric double traceless
tensors $\Phi_{M_1\dots M_J}$. The corresponding action is
invariant under the gauge transformations $\delta\Phi_{M_1\dots
M_J}=\nabla_{(M_1}\xi_{M_2\dots M_J)}$, where $\nabla$ denotes the
covariant derivative with respect to the general coordinate
transformations and the gauge parameter $\xi$ is a traceless
symmetric tensor. The action for free HSF in the AdS$_5$ space
reads
\begin{equation}
\label{23}
S^{(J)}=\frac{1}{2}\int d^4\!x\, dz \sqrt{g}\,
e^{-az^2}\left(\nabla_N\Phi_{M_1\dots M_J}\nabla^N\Phi^{M_1\dots
M_J}+\dots\right),
\end{equation}
where further terms are omitted. In the gauge $\Phi_{z\dots}=0$,
the action for a rescaled field
\begin{equation}
\label{res}
\Phi=\left(\frac{z}{R}\right)^{2(1-J)}\tilde{\Phi},
\end{equation}
contains only the first kinetic term displayed in~\eqref{23}. The
equation of motion for $\tilde{\Phi}(x)$ results in the mass
spectrum~\cite{son2}
\begin{equation}
\label{24}
m_{n,J}^2=4a(n+J),\qquad n=0,1,2,\dots.
\end{equation}
Note that $a>0$ in spectrum~\eqref{24}. For $a<0$ the spectrum is
given by relation~\eqref{6} for any $J$. This was another reason
in favor of the choice $a>0$ in Ref.~\cite{son2}. In the framework
of a more general description of HSF, the restriction $a>0$ is not
necessary~\cite{br4}.

The rescaled field $\tilde{\Phi}_{M_1\dots M_J}$ corresponds to a
twist-two operator with the canonical dimension
$\Delta=J+2$~\cite{son2}. Such operators have the lowest dimension
for given spin and play the dominant role in interpolating the
hadron states. It is interesting to observe that
action~\eqref{23} for rescaled fields $\tilde{\Phi}$ can be then
written in a compact form using relation~\eqref{3a},
\begin{equation}
\label{26}
S^{(J)}=\frac{1}{2}\int d^4\!x\, dz
\sqrt{g}\left(\frac{R}{z}\right)^{R^2m_5^2}\!
e^{-az^2}\nabla_N\tilde{\Phi}_{M_1\dots
M_J}\nabla^N\tilde{\Phi}^{M_1\dots M_J},
\end{equation}
where $R^2m_5^2=4(J-1)$.

\section{A solvable extension of SW model to arbitrary intercept}

The linear Regge spectrum~\eqref{24} reproduces the prediction of
Veneziano amplitude and the spectral law $m_{n,J}^2 \sim n+J$
seems to agree with available experimental data on the light
non-strange mesons~\cite{epj}. However, the intercept of real
Regge trajectories differs from that of~\eqref{24},
\begin{equation}
\label{24b}
m_{n,J}^2=4a(n+J+b),\qquad n=0,1,2,\dots,
\end{equation}
where $b$ should be a phenomenological parameter regulating the
phenomenological intercept. The question appears, how to reproduce
spectrum~\eqref{24b} in simple solvable SW models? A seemingly
straightforward idea is to change the 5D masses. However, these
masses are already fixed by the requirement that we interpolate
the meson states by the quark currents of lowest canonical
dimension. It looks more reasonable to modify the dilaton
background in the SW action. A method for derivation of modified
background that leads to spectrum~\eqref{24b} in the vector case
$J=1$ was proposed in Ref.~\cite{genSW}. This method can be
extended to arbitrary spin. Below we give the final answer for
$a>0$ that generalizes expression~\eqref{26},
\begin{equation}
\label{25}
S^{(J)}=\frac{1}{2}\int d^4\!x\,dz\sqrt{g}
\left(\frac{R}{z}\right)^{R^2m_5^2}U^2(b,-|J-1|;az^2)e^{-az^2}\mathcal{L}^{(J)},
\end{equation}
where $U$ is the Tricomi hypergeometric function, $J=0,1,2,\dots$,
and we must set $R^2m_5^2=2$ in the case $J=0$. The HSF in the
Lagrangian $\mathcal{L}^{(J)}$ are implied in the rescaled
form~\eqref{res} as in~\eqref{26}. It should be noted that the
background in~\eqref{23} is universal for any spin $J$. However,
in terms of "physical" rescaled fields~\eqref{res} it becomes
spin-dependent as can be seen in~\eqref{26}. The generalized
background in action~\eqref{25} is also spin-dependent.

\section{On the chiral symmetry breaking}

In the presented approach, the hadron masses depend on the
dimension $\Delta$ and spin $J$ of the corresponding interpolating
operator. This leads to degeneracy of states with opposite space
parity when they are described by operators with equal $\Delta$
and $J$, e.g. the $\rho$ and $a_1$ mesons. It is commonly accepted
that the observable large mass splittings between such states are
caused by the Chiral Symmetry Breaking (CSB) in QCD. The Hard-wall
model incorporates the CSB in much the same way as traditional
chiral effective theories~\cite{son1}. Such a description of the
CSB dynamics does not work, however, in simple SW
models~\cite{son2}. The description of CSB in Ref.~\cite{son1}
requires introduction of cubic, quartic and higher vertices which
give rise to meson decays and interactions. As the CSB effect is
(most likely) not suppressed by large number of colors $N$, it is
difficult to adjust such a description with the large-$N$ limit,
in which the holographic duality is formulated. In addition, this
description does not provide the mass splittings between parity
partners among higher spin mesons in which apparently the
splittings are of the same order as between $\rho$ and $a_1$
mesons~\cite{epj}.

In my opinion, a self-consistent holographic description of CSB
should be based on quadratic in fields 5D actions. But this is an
open problem. A solution might lie in reinterpretation of what we
call spin in the holographic models. An interesting possibility
was put forward by the light-front holographic QCD~\cite{br4}
where, similarly to the non-relativistic potential models, the
mass degeneracy of parity partners is lifted due to different
orbital momentum of quark constituents. An alternative option
could consist in interpretation of intercept parameter $b$ in
spectrum~\eqref{24b}: As was advocated in Ref.~\cite{genSW} for
vector mesons, the sign of $b$ depends on parity.

\section{Conclusion}

In summary, the bottom-up holographic approach is an interesting
and mathematically nice method for description of excited hadron
spectrum even if in reality it is not rigorously related with some
underlying string theory.

\section*{Acknowledgments}

The work was supported by the Saint Petersburg State University
research grant 11.38.189.2014 and travel grant 11.41.601.2016, and
also by the RFBR grant 16-02-00348-a.


\end{document}